# Quasi ballistic magnetization reversal

*H.W. Schumacher[1], C. Chappert[1], R.C. Sousa[2], P.P. Freitas[2], J. Miltat[3]*

[1]  Institut d'Electronique Fondamentale, UMR 8622, CNRS, Université Paris Sud, Bât. 220, 91405 Orsay, France

[2]  Instituto de Engenharia de Sistemas e Computadores, Rua Alves Redol, 9, 1° Dt., P-1000 Lisboa, Portugal

[3]  Laboratoire de Physique des Solides, UMR 8502, CNRS, Université Paris Sud, Bât. 510, 91405 Orsay, France

**Abstract**

We demonstrate a quasi ballistic switching of the magnetization in a microscopic magneto resistive memory cell. By means of time resolved magneto transport we follow the large angle precession of the free layer magnetization of a spin valve cell upon application of transverse magnetic field pulses. Stopping the field pulse after a 180° precession rotation leads to magnetization reversal with reversal times as short as 165 ps. This switching mode represents the fundamental ultra fast limit of field induced magnetization reversal.

**Corresponding Author:**

Hans Werner Schumacher, IEF, Université Paris Sud, Bât. 220, F-91405 Orsay, France,

e-mail: schumach@ief.u-psud.fr, phone: +33169154013, fax: +33169154000

**Article**

The fundamental ultra fast limit of field induced magnetization reversal in ferromagnets is directly related to the precession frequency of the magnetization (1) upon application of a magnetic field pulse. When applying a field oriented mainly antiparallel to the initial magnetization *M*, *M* has to undergo multiple precessional oscillations about the local effective field to reach full alignment with the reversed equilibrium direction (2,3,4). The resulting reversal times are thus considerably longer than one precession period and are generally of the order of nanoseconds. A novel approach towards ultra fast magnetization reversal is the so-called precessional switching of magnetization (5,6,7,8,9). Here, application of fast rising field pulses perpendicular to the initial direction of *M* initiates a large angle precession (10,11,12) that is used to reverse the magnetization. The ultimate switching speed could then be reached by stopping the field pulse exactly after a 180° precessional rotation (7,8). This way, magnetization reversal in microscopic magnetic memory cells by field pulses as short as 140 ps has recently been demonstrated (13,14). However, due to the lack of time resolution in these experiments the effective reversal times are yet unknown and could be limited to several nanoseconds by the decay time of residual magnetic precession ("ringing") upon field pulse termination (10,11,12). According to theoretical predictions (8), however, an exact control of the pulse parameters should allow to switch the magnetization on so-called ballistic trajectories characterized by the absence of ringing and thus to reach the fundamental limit of reversal speed.

In this letter, we experimentally explore the fundamental limit of ultra fast magnetization reversal times in a microscopic memory cell. By measuring the time resolved magneto resistance response of a spin valve during application of transverse field pulses we follow the pronounced precession of the cell's free layer magnetization. Stopping the field pulse exactly after a half precessional rotation induces a quasi ballistic reversal characterized by switching times as short as 165 ps and, within measurement accuracy, by a complete suppression of long wavelength magnetic excitations after field pulse decay.

An optical micrograph of a microscopic magnetic cell used in our experiments is shown in Fig. 1(a). The stadium shaped spin valve (SV) of 5 µm x 2.3 µm lateral dimensions consists of Ta 65Å / NiFe 40 Å / MnIr 80 Å / CoFe 43 Å / Cu 24 Å / CoFe 20 Å / NiFe 30 Å / Ta 8 Å. The exchange bias field defining the direction of the pinned magnetic layer and the magnetic easy axis of the free layer are oriented along the long dimension of the cell. The electrical contacts (C1,C2) allow to measure the cell's giant magneto resistance and thus to derive the average angle between the magnetization of the free and of the pinned layer (15). Due to the overlap between the contact pads and the SV the measured magneto resistance (MR) is dominated by the magnetization orientation in the 2.1 µm wide center region not covered by the two contacts. The field pulses $H_{Pulse}$ are generated by current pulse injection into a buried pulse line (*PL*). All electrical lines are integrated into high bandwidth coplanar waveguides. Flowing a DC current through the SV during pulse application allows to measure the SV's ultra fast MR response using a 50 GHz sampling oscilloscope (16). The current pulses are characterized after transmission through the device using a second oscilloscope channel. As sketched in Fig 1(b) the field pulse $H_{Pulse}$ is applied along the in-plane magnetic hard axis i.e. perpendicular to the initial and final direction of *M* whereas external static fields $H_{Stat}$ are applied along the easy axis by an external coil. The maximum pulsed fields obtained on this device are of the order of 280 Oe with pulse durations down to 170 ps and rise times as short as 45 ps (10 to 90 % amplitude variation). For the sampling measurements the pulses are applied with a repetition rate of 5 KHz. 2 µs after the fast pulse a 140 ns, 80 Oe reset pulse generated by a second macroscopic field line located on the back of the sample resets $M$ to the initial state $M_i$. By averaging over several hundred curves low noise levels of the order of 50 µV are obtained (16) corresponding to an angular resolution down to 1.5°.

A static *MR* hysteresis loop of the SV is shown in Fig 2(a). The coercivity $H_C$ is 20 Oe and the loop is shifted to an offset field of $H_{Offset}$ = 20 Oe (17). In the following, this offset field is always compensated by an external static field $H_{Stat} = H_{Offset}$. In the loop center i.e. at offset compensation the measured *MR* change due to the reversal of the free layer is of the order of 4%.

Fig 2(b) shows the time resolved *MR* response to hard axis field *steps* of amplitudes $H_{Pulse}$ between 53 and 272 Oe. The amplitude is varied in 1 dB steps. Pronounced oscillations of the MR are a clear indication of magnetic precession. From the exponential decay of the precession amplitude with time (8) we derive an effective damping parameter $\alpha = 0.03 \pm 0.005$. From Fourier transformation of the *M*R response we obtain $f_{Prec}$, the precession frequency under hard axis excitation, plotted in Fig. 2(c) vs. field amplitude $H_{Pulse}$. Kittel's formula for ferromagnetic resonance (18) captures the basic physics of the observed magnetization motion: the cell free layer is modeled as an ellipsoid with demagnetizing factors $N_X/4\pi = 0.0005$ (easy axis), $N_Y/4\pi = 0.00217$ (in-plane hard axis), and $N_Z/4\pi = 0.99733$ (out of plane) as suitable for the free layer geometry. Furthermore, the usual saturation magnetization $4\pi M_S = 10800$ Oe is assumed for permalloy. The calculated dependence of $f_{Prec}(H_{Pulse})$ given by the gray straight line is in good agreement with the calculated resonance frequencies of the SV's free layer. The MR signal is clearly dominated by the free layer precession and not by the dynamics of the pinned layer of the SV (19). In the following interpretation of the time resolved data we thus neglect the pinned layer dynamics and directly calculate $m_X$ the component of $M$ along the easy axis from the measured MR signal (15,20).

To model the time dependence of the magnetization response to the ultra short hard axis pulses we solve the Landau-Lifshitz-Gilbert equation (1)

$$\frac{dM}{dt} = -\gamma (M \times H_{eff}) + \frac{\alpha}{M_S}\left(M \times \frac{dM}{dt}\right)$$

in the single spin (or macro spin) approximation. Here, $\gamma$ is the gyromagnetic ratio, and $\alpha$, $M_S$, the damping parameter and the saturation magnetization, respectively. The free layer is again modeled using the demagnetizing factors, saturation magnetization, and damping parameter given above. Furthermore, the pulse parameters used in the simulations mimic the shape and amplitude of the experimental pulses.

The simulated magnetization response to a 81 Oe field step along the in-plane hard axis is displayed in Fig. 2(d), showing the projection of the trajectory of $M$ on the $x$-$y$-plane. $m_X$ is the normalized component of $M$ along the easy axis and $m_Z$ the out-of-plane component. $M$ responds to the field step by damped precession about $H_{Pulse}$ starting from its initial position $M_i$ (aligned along $-m_X$) and ending up being aligned with the hard axis field. Due to the sample's shape anisotropy and the resulting strong demagnetizing fields $M$ remains mainly in plane (notice the different scales for $m_X$ and $m_Z$).

In Fig. 3(a,d) two calculated magnetization responses to short *pulses* are presented in the same way. The pulse in (a) has an amplitude of $H_{Pulse} = 81$ Oe and a duration of $T_{Pulse} = 175$ ps. Its time evolution is displayed in Fig. 3 (b) (black line). During pulse ap-plication, $M$ performs approximately a half precession turn about $H_{Pulse}$ (see 3(a)). At pulse decay time, $M$ is oriented near the reversed easy axis direction $M_f = -M_i$ and relaxes towards it. The hard axis pulse is thus expected to induce a precessional switching of the magnetization. The measured field pulse (gray) is displayed in Fig. 3(b) whereas the measured (gray) and calculated (black) magnetization responses are given in Fig. 3(c) where $m_X$ is plotted as a function of time. The measured time evolution of $m_X$ is well described by the simple simulation. The short pulse induces precessional switching of the magnetization with ultra short measured reversal time ($-0.9m_Z$ to $+0.9m_Z$) $T_{Switch} = 165$ ps. Furthermore, no significant precession after pulse decay is present in $m_X$ neither in the measured data nor in the simulation and long wavelength magnetic excitations after pulse decay ("ringing") are suppressed. The pulse very nearly matches the half precessional turn $T_{Pulse} \approx \frac{1}{2} \cdot T_{Prec}$ and, thus, the magnetization switches quasi ballistically (8) i.e. with a close to optimum trajectory towards the reversed direction. This quasi ballistic switching represents the fundamental ultra fast limit of field induced magnetization reversal for the given field amplitude.

The trajectory in Fig. 3(d) is the calculated response to a 205 Oe, 240 ps pulse. Due to the higher field and longer pulse duration time $M$ now performs a full 360° rotation about $H_{Pulse}$ before the pulse decays. Upon pulse termination, $M$ is in this case oriented near the initial easy direction $M_i$ and relaxes towards the latter. Therefore, in spite of strong precession during pulse application no effective magnetization reversal takes place. The corresponding pulses and magnetization responses are found in Fig. 3(e,f), respectively. Now, during pulse application, $m_X$ oscillates from the initial direction to the reversed orientation and back. The pulse matches here a full precessional rotation $T_{\text{Pulse}} \approx T_{\text{Prec}}$. Again, only little ringing is found following pulse termination.

For weak damping and sufficiently high field strengths (7, 14) higher ratios of $T_{\text{Pulse}}/T_{\text{Prec}}$ can reveal consecutive regions of precessional higher order switching and non switching. Switching then occurs whenever $T_{\text{Pulse}} \approx (n+\tfrac{1}{2}) \cdot T_{\text{Prec}}$ with n = 0,1,2,... being the order of the switching event. In contradistinction, no effective cell reversal will occur whenever $T_{\text{Pulse}} \approx n \cdot T_{\text{Prec}}$. This effect can be clarified using the macro spin trajectory in Fig. 2(d). Stopping the pulse at a point of the precession trajectory with a positive value of $m_X$ (i.e. on the right hand side of the dashed vertical line $m_X = 0$) will lead to relaxation into the reversed magnetization state (switch) whereas pulse termination at $m_X < 0$ will result into relaxation back to the initial state (no switch).

This oscillatory nature of precessional switching by hard axis pulses is well observed in Fig. 4. In (a), the measured response of $m_X$ to a 305 ps, 272 Oe pulse is plotted as a function of time. In (b) the response of $m_X$ to a series of pulses with $H_{\text{Pulse}} = 272$ Oe (21) and $T_{\text{Pulse}} = 200,...,900$ ps (10 ps increments) is plotted as a gray scale map, as a function of time and pulse duration. The curve in (a) is a section through the data in (b) along the horizontal dashed line. As seen in (a), the 305 ps pulse induces a first order ($n$=1) precessional switch ($T_{\text{Pulse}} \approx 1.5 \cdot T_{\text{Prec}}$). $M$ rotates one and a half times about $H_{\text{Pulse}}$ during pulse application corresponding to a triple change of sign of $m_X$. However, the pulse parameters are not well tailored to 1.5 precessional rotations. As a consequence, $M$ is not fully aligned with the reversed easy direction upon pulse termination resulting in a residual precession upon pulse termination (see arrows (1)). For the given pulse, full alignment of $M$ with the final easy axis direction takes more than 1 ns. The precession limited effective reversal time is thus considerably longer than the switching pulse duration of only 305 ps. This underlines the importance of a precise control of the pulse parameters to achieve ultra fast quasi ballistic switching.

The multiple oscillations of $M$ about $H_{Pulse}$ for longer pulses can also be seen in Fig. 4(b). The oscillation maxima of $m_X$ (light regions running vertically) are marked by the black dots and numbers 0-3 on the upper border of the gray scale plot. The pulse duration is indicated in the data by the inclined dotted line. Pulse decay at a maximum of $m_X$ (i.e. with $M$ oriented near the reversed easy direction) inevitably leads to relaxation to the reversed easy axis direction i.e. to high order precessional switching (white horizontal regions after pulse termination with switching order $n$). On the contrary, pulse decay at a minimum of $m_X$ (i.e. with $M$ oriented near the initial easy direction) always leads to relaxation towards the initial direction of $M$ (gray horizontal regions) and no effective switching is monitored. Again, near the transitions from switching to non-switching the alignment of $M$ and the final easy axis direction is poor and a pronounced ringing of the magnetization upon pulse decay occurs (see e.g. arrows (2)).

Concluding, we have experimentally reached the fundamental ultra fast limit of field induced magnetization reversal of a microscopic magnetic memory cell. Quasi ballistic magnetization switching with ultra short reversal times of only 165 ps was demonstrated. The moderate field strength of 81 Oe and the low switch pulse energy of only 27 pJ for the present device demonstrates a high efficiency when compared to standard magnetization reversal schemes (2,3). Such ballistic switching could e.g. open the door to ultra fast and yet low power magnetic random access memories (M-RAMs) (22,23) with clock rates well above the GHz. (24)

**FIGURE CAPTIONS:**

**FIGURE 1:**

Magnetic memory cell used in the precessional switching experiments. (a) optical micrograph. The stadium shaped spin valve cell (SV) is located in the center of the image. The ends are covered by the two electrical contacts (C1, C2, surrounded by the dotted lines). A 4 µm wide buried pulse line (PL, marked by the white dashed line) is aligned with the long axis of the SV in order to create magnetic field pulses oriented along the in-plane hard (short) axis of the cell. (b) sketch of the magnetic field configuration: The pulse field $H_{Pulse}$ (along $y$) is applied perpendicular to the initial and final magnetization $M_i$, $M_f$. External static fields $H_{Stat}$ (along $x$) are applied along the easy (long) axis.

**FIGURE 2:**

(a) easy axis magneto resistance (MR) hysteresis loop of the spin valve cell. Magnetization reversal of the SV's free layer leads to a MR change of ~ 4 %. The loop is shifted to an offset field of $H_{Offset} = 20$ Oe. (b) time resolved MR response of the SV cells to magnetic hard axis field steps ranging from 53 Oe (bottom) to 272 Oe (top) in 1 dB increments. The MR data is offset for clarity. Precessional oscillations of the free layer about $H_{Pulse}$ are clearly observed. (c) magnetic precession frequency $f_{Prec}$ about the applied field vs. field amplitude as derived from Fourier transforms of the data in (b). The frequency-dependence is well described by ferromagnetic resonance of the free layer (gray line). (d) calculated trajectory of the magnetization $M$ in response to a field step of $H_{Pulse} = 81$ Oe. $m_X$, $m_Z$ are the normalized components of $M$ along the easy axis, and out of plane, respectively (cp. Fig. 1(b)).

**FIGURE 3:**

Precessional switching of a SV cell. Calculated trajectories of precessional switching (a) and non-switching (d) in the $m_X$-$m_Z$ plane. For switching the field pulse is stopped after a 180° precessional turn about $H_{Pulse}$ (a). ($H_{Pulse}$ = 81 Oe, $T_{Pulse}$ = 175 ps). Higher fields and longer pulse durations ($H_{Pulse}$ = 205 Oe, $T_{Pulse}$ = 240 ps) induce a full 360° precessional rotation (d) and *M* relaxes towards the initial easy axis direction after pulse decay (no switch). Field (b) and magnetization component $m_X$ (c) *vs.* time for the $H_{Pulse}$ = 81 Oe, $T_{Pulse}$ = 175 ps pulse. Gray dots: experiment, black lines: simulation. $m_X$ switches within $T_{Switch}$ = 165 ps. After pulse decay no residual precession is found indicating optimum quasi ballistic reversal. Field (e) and $m_X$ (f) *vs.* time for the $H_{Pulse}$ = 205 Oe, $T_{Pulse}$ = 240 ps pulse. $m_X$ toggles towards the reversed direction and back during pulse application: *M* is not reversed upon pulse termination.

**FIGURE 4:**

Higher order precessional switching: (a) First order switch. Measured $m_X$ *vs.* time for $H_{Pulse}$ = 272 Oe, $T_{Pulse}$ = 305 ps. After pulse decay residual precession occurs (arrows (1)). (b) gray scale encoded map of $m_X$ as a function of time and pulse duration $T_{Pulse}$. White: $m_X$ =1, dark gray, $m_X$ = -1. Nominal pulse amplitude $H_{Pulse}$ = 272 Oe (21). Pulse field decays to zero along the inclined dotted line. (a) corresponds to a section through (b) along the dashed horizontal line. Higher order switching (white horizontal regions, switching order *n* is indicated) occurs in phase with the precession at pulse cut-off. Zero order switching (*n*=0) is not accessible for the given pulse amplitude.

(24) HWS acknowledges financial support by the European Union (EU) Marie Curie fellowship HPMFCT-2000-00540. The work was supported in part by the EU Training and Mobility of Researchers Program under Contract ERBFMRX-CT97-0147, and by a NEDO contract "Nanopatterned Magnets".

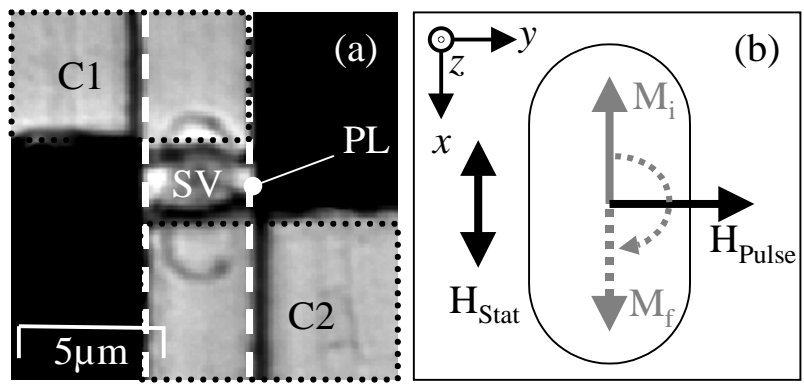

Figure 1
H.W. Schumacher et al.

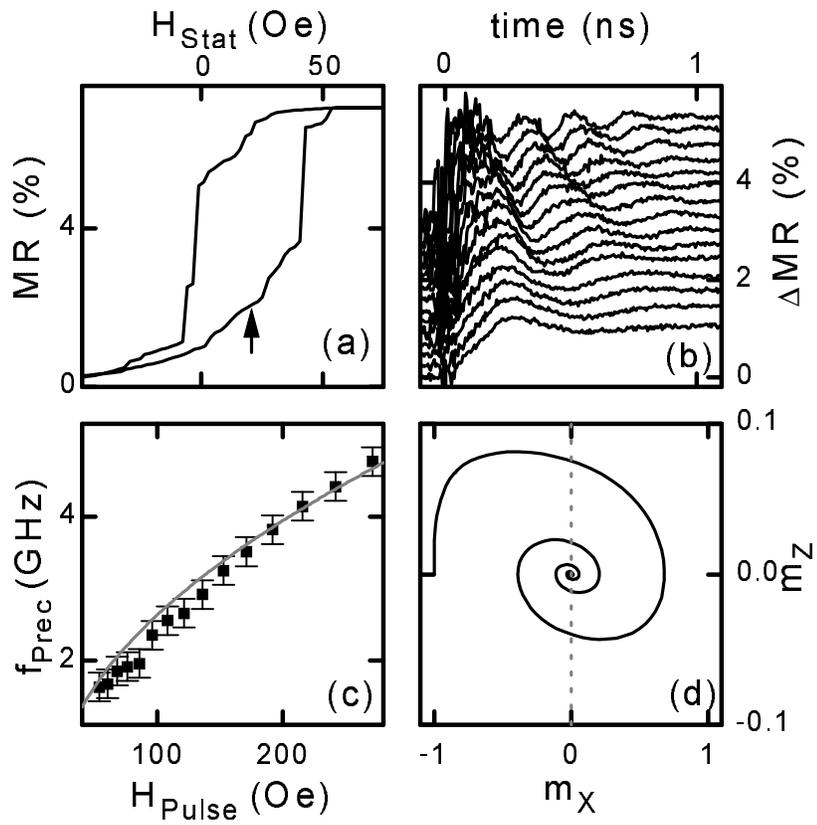

Figure 2

H.W. Schumacher et al

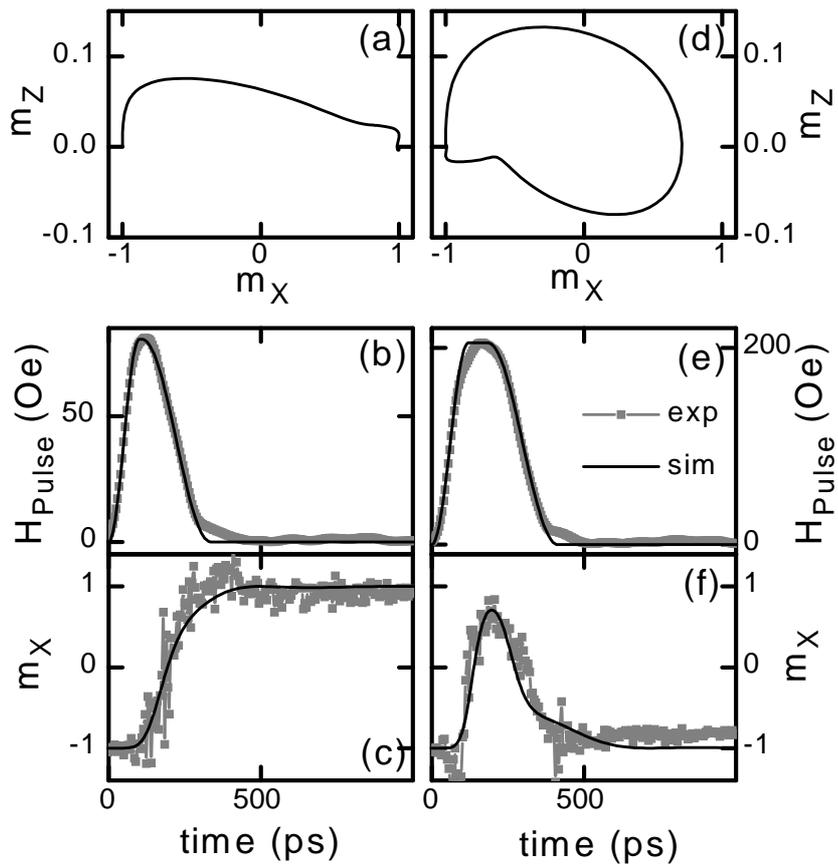

Figure 3

H. W. Schumacher et al

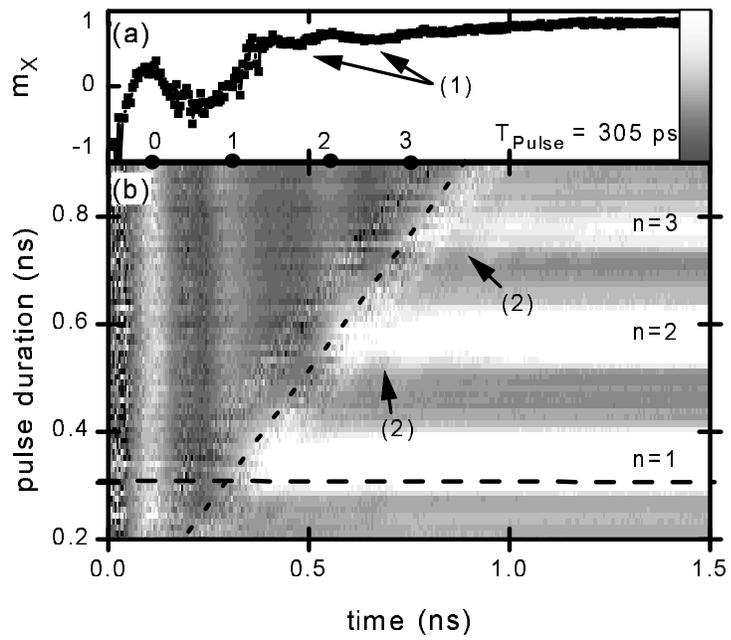

Figure 4

H.W. Schumacher et al.